\documentclass[review]{elsarticle}
\usepackage{natbib}
\usepackage{geometry}
\usepackage{fleqn}
\usepackage{graphicx}
\usepackage{txfonts}
\usepackage{textcomp}
\usepackage{subcaption}
\usepackage{multirow}
\usepackage{xcolor}
\usepackage[normalem]{ulem}
\graphicspath{{epsFigures_reducedResolution/}}
\biboptions{sort&compress}
\usepackage{hyperref}

\begin{document}
	
	\title{Shadowing effect on the morphologies of thin-films 
		deposited over plane and patterned substrates}
	\author[1]{S. Bukkuru \corref{cor1}}
	\ead{srinivasaraobukkuru@gmail.com}
	\author[2]{H. Hemani}
	\ead{harshscience777@gmail.com}
	\author[3]{S. Maidul Haque}
	\ead{maidul@barc.gov.in}	
	\author[1,4]{M. Ranjan}
	\ead{ranjanm@ipr.res.in}
	\author[1]{J. Alphonsa}
	\ead{alphonsa@ipr.res.in}
	\author[3,4]{K. Divakar Rao}
	\ead{divakar@barc.gov.in}
	\author[2,4]{M. Warrier}
	\ead{manoj.warrier@gmail.com}
	
	\cortext[cor1]{Corresponding author}
	\address[1]{Facilitation Centre for Industrial Plasma Technologies,
		IPR, Gandhinagar, Gujarat, India - 382 016}
	\address[2]{Computational Analysis Division, BARC, Visakhapatnam,
		Andhra Pradesh, India - 531 011}
	\address[3]{Atomic \& Molecular Physics Division, BARC, Visakhapatnam,
	    India - 531 011}
	\address[4]{Homi Bhabha National Institute, Anushaktinagar, Mumbai, 
		Maharashtra, India - 400 094}
	
	\begin{abstract}
		A simple two-dimensional ballistic deposition (2D-BD) code has
		been developed to show the geometric shadowing effects in
		thin-film evolution. A criterion to determine when the
		shadowing effect dominates over surface diffusion is presented.
		This code is validated for a plane substrate by comparing the
		morphological features like the angle of growth, porosity and
		root mean square (RMS) surface roughness of simulated
		thin-film deposits with published results.  The code is then
		further validated by applying to two experiments of thin film
		deposition on (i) a plane substrate with a parallel collimator
		and (ii) a sinusoidal patterned substrate. In the case of a
		plane substrate with a parallel collimator, the reported
		thin-film morphologies ranging from continuous films to
		nano-islands have been reproduced. For a patterned substrate,
		the experimentally observed replicability of a substrate
		topography by the surface of thin-film for normal and oblique
		angle deposition is ascertained. The code predicts an angle of
		growth of 56$\pm$5\textdegree{} for an angle of deposition of
		77\textdegree{} on patterned substrates, compared to the
		experiment which reports 50\textdegree{}. This is much better
		than empirical rules like the tangent rule, which predicts
		71\textdegree{} for this case.
	\end{abstract}
	
	
	\maketitle
	
	\section{Introduction}
	The thin-film deposition was reported
	more than 150 years ago by W. R. Grove \cite{grove1852vii}
	and M. Faraday \cite{faraday1857x}. However, a large
	variety of applications of thin-films in optical, sensor and energy
	storage devices have been witnessed in the
	past few decades \cite{hawkeye2014glancing,nikitenkov2017modern,lakhtakia2005sculptured,
	mattox2003vacuum}. This became possible with the advancement of its
	subclasses such as “Oblique Angle Deposition (OAD)'',
	``Glancing Angle Deposition (GLAD)”, etc. \cite{martin2009handbook}. 
	OAD is performed typically with a fixed substrate, inclined at any 
	angle to the incoming flux of depositing particles. Whereas in GLAD,
	which is an extension of OAD, the substrate is fixed at glancing angles
	that can result in nano-structures, such as inclined, helical or 
	zig-zag columns, etc. \cite{zhao2003designing}. In GLAD, the substrate
	may also undergo individual or combined effects of change in relative
	position, inclination or rotation. Use of
	a collimator or a patterned substrate gives additional freedom to
	tune the thin-film properties
	\cite{Divakar,keller2011polycrystalline}.		
	% subject to the criterion being satisfied
	
	Ballistic deposition simulations are used to understand the
	critical growth mechanisms of thin-film evolution and to predict
	their morphologies \cite{review2016}. In ballistic deposition (BD),
	depositing particles are represented by hard spheres which are
	projected towards a substrate at a chosen angle depending on the
	source location. The particles are projected one after the other.
	Each projected particle travels linearly until it either reaches the
	substrate or it intercepts a previously deposited particle. It is
	assumed to stick to a particle or substrate with which it makes
	first contact. During such a deposition, a projected particle can
	be shadowed by the previously deposited particles. This is called 
	the \textit{shadowing effect}
	\cite{henderson1974simulation,dirks1977columnar,karabacak2011thin}
	in thin-film deposition. The initial ballistic deposition
	simulations of Henderson et al. \cite{henderson1974simulation}
	provide details about the columnar morphology in thin-films.
	Later, \textit{bouncing} and \textit{relaxation} of projected particles
	\cite{kim1977computer}, \textit{surface diffusion} and
	\textit{re-emission} \cite{jones1967re,drotar2000mechanisms,
	drotar2000surface} of deposited particles have been introduced
	depending on the experimental growth parameters like temperature,
	energy of the incident	particle, etc. Meakin et al. have carried out
	extensive 2D and 3D ballistic deposition simulations using fractals. They show that the differences in the angle of growth for the 2D and 3D simulations is 3--6\textdegree{}  for angles of 
	deposition 60--85\textdegree{} \cite{meakin1988ballistic,meakin1998fractals}. Ballistic deposition
	simulations are computationally inexpensive as compared to molecular
	dynamics (MD) simulations and can simulate much larger sizes of deposits
	to study the experimentally observable morphological features.
		
	A \textit{two-dimensional ballistic deposition} code (2D-BD), based
	on geometric shadowing effect to study the morphological features
	in thin-film evolution has been developed and it is described in 
	Section \ref{Method}. Surface diffusion of the deposited particles 
	is not considered. A criterion for neglecting surface diffusion
	considering the incident fluxes and surface migration energy of 
	deposited particles is presented. The code is validated with
	published results of angle of growth, porosity and RMS surface
	roughness for a plane substrates in Section \ref{Results-2DBD}. It is
	then further validated by applying it to two experiments
	in Sections \ref{Results-CGLAD}, \ref{Results-patternedSub}. One
	experiment \cite{Divakar} demonstrates the tuning of the thin-film
	morphology using a novel angle constrained glancing angle
	deposition on a plane substrate with a parallel collimator. The
	other experiment \cite{keller2011polycrystalline} shows the height
	to which a patterned substrate topography is replicated by the film
	surface under normal and oblique angle depositions. Finally, the
	conclusions are summarized in Section \ref{Conclusions}.
	
	\section{Simulation Method}
	\label{Method}	
	The present study is limited for cases where the typical time for
	surface diffusion of a deposited particle is much less than the
	typical time for deposition of incident particles on it. If
	$\nu_{dep}$ is frequency of incident particles depositing on
	previously deposited particles and $\nu_{dif}$ is the frequency of
	the deposited particle making a random jump (diffusion), then,
	
	\begin{equation}
	\nu_{dep} = \Gamma_{i} \times \sigma_p 
	\end{equation}
	\begin{equation}
	\nu_{dif} = \omega_0 \times e^{(\frac{-E_m}{k_BT})}
	\end{equation}
	
	\noindent
	Here, $\Gamma_{i}$ is the flux of incident particles onto the
	target and $\sigma_p$ is the cross-sectional area of a deposited
	particle for an incident particle. The deposited particles diffuse
	with a typical jump-attempt frequency or phonon frequency
	($\omega_0$) of around $10^{12} s^{-1}$. $E_m$ is the migration
	energy required for surface diffusion of a deposited particle,
	$k_B$ is the Boltzmann's constant and T is temperature. The 2D-BD
	simulations are therefore valid for the following criterion:
	\begin{equation}
	\frac{\nu_{dep}}{\nu_{dif}} > 1 \hspace{0.5cm}
	\texttt{or} \hspace{0.5cm}
	\Gamma_{i} > \frac{\omega_0 \times e^{(\frac{-E_m}{k_BT})}}{\sigma_p}
	\end{equation}
	
	The migration energy of a deposited particle depends on the nature
	of adsorption. In the case of physisorption, adsorbate
	particles stick with a substrate due to weak Van der Waal forces
	and have migration energies of the order of 0.1 eV. Chemisorption
	involves chemical interaction between adsorbate particles and the
	substrate and has migration energies of the order of 0.9 eV. In
	case the deposited atoms form chemical bonds (ionic, covalent or
	metallic) with a substrate, the migration energies can be of the
	order of 1.5 eV. Assuming the above stated typical values of
	migration energy, a $\sigma_p$ of 1.5 \AA, a substrate temperature
	of 70~\textcelsius{} and typical phonon frequency of $10^{12}
	s^{-1}$, the 2D-BD code will be valid for incident fluxes greater
	than $10^{29}$ $m^{-2} s^{-1}$ for physisorption, $10^{17}$ $m^{-2}
	s^{-1}$ for chemisorption and $10^9$ $m^{-2} s^{-1}$ for
	ionic/covalent/metallic bonds. Note that the temperature of the
	substrate is also an important parameter in the calculation of the
	criterion for applicability of the code. 2D-BD simulations are
	carried out to study the thin-film deposition on (i) a plane
	substrate, (ii) a plane substrate with a parallel collimator and
	(iii) a sinusoidal patterned substrate. These simulation methods
	are described below.
	
	In the present study, circular discs of a fixed size are used to
	represent the particles. These are initialised uniformly, randomly
	and then projected towards the substrate at a chosen angle of
	deposition. This angle is chosen by sampling from a Gaussian
	distribution with the given angle of deposition ($\alpha$) and a
	standard deviation ($\sigma$). Each particle is projected after
	the deposition of a previously projected particle. A projected 
	particle travels rectilinearly and reaches the 
	substrate or a previously deposited particle. It is assumed to stick
	to a particle or substrate with which it makes its first contact.
	
	\begin{figure}
		\centering	
		\includegraphics[width=0.7\linewidth]{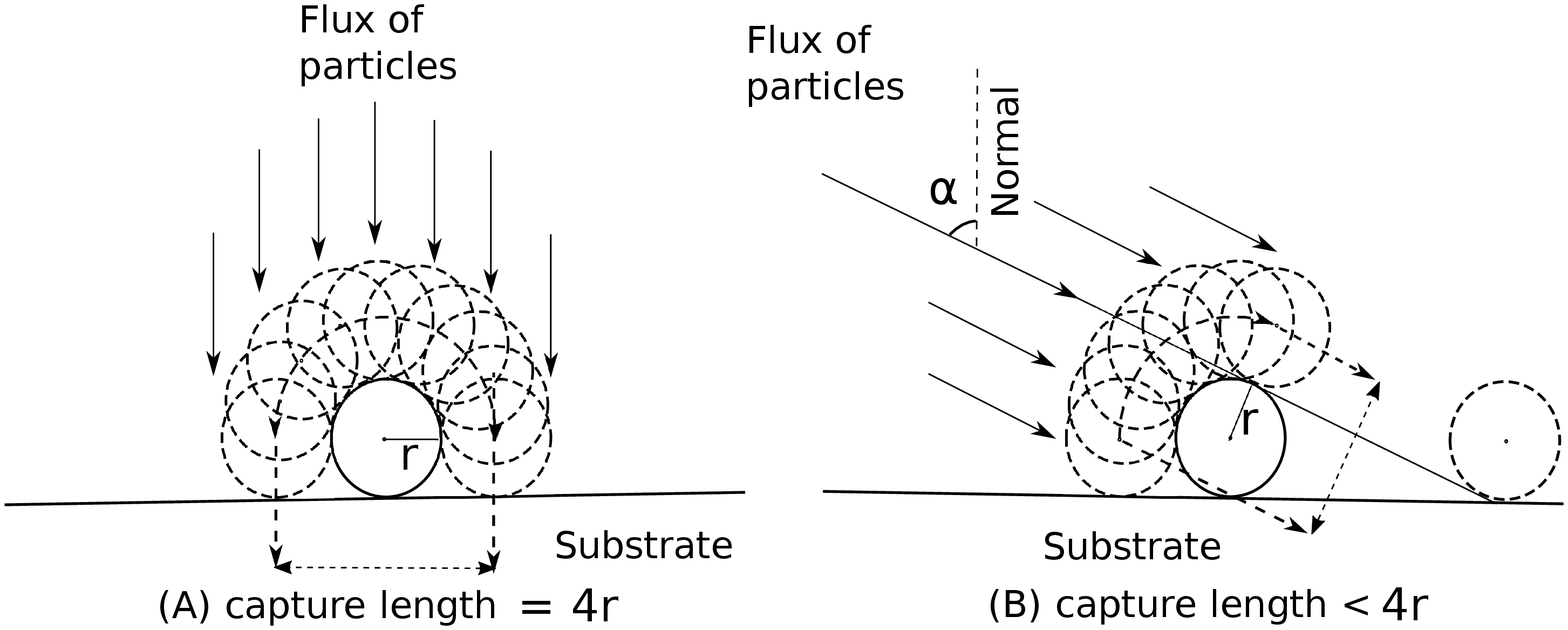}
		\caption{In normal deposition, a projected particle can undergo
			symmetric impingement about a previously deposited particle.
			Due to the asymmetric impingement in oblique angle deposition,
			the capture length reduces and some portion on the substrate is shadowed.}
		\label{capture_radius}
	\end{figure}
	
	During the normal ballistic deposition, a projected particle can
	undergo symmetric impingement about a previously deposited particle
	as shown in Fig \ref{capture_radius} (A). This results in dense
	packing of the particles. In oblique angle deposition, a particle
	deposited earlier shadows some portion of the deposition area  as
	shown in Fig \ref{capture_radius} (B). Due to this, the capture
	length of particles arriving later will be reduced by the particles
	that have previously been deposited, resulting in the
	\textit{shadowing effect} \cite{henderson1974simulation,
	dirks1977columnar}. With this asymmetric impingement, the mean
	orientation of the pair of particles shifts towards the normal.
	This makes the angle of growth ($\beta$) of a thin-film less than
	the angle of deposition ($\alpha$) \cite{dirks1977columnar}.
		
	\subsection{2D Ballistic Deposition on a plane substrate}
	\label{Method-2DBD}
	Particles are initialized uniformly, randomly and then projected
	towards the substrate at a chosen angle as described before. 2D-BD
	simulations have been carried out for the angles of deposition
	$20$\textdegree{} - $80$\textdegree{} in steps of
	$10$\textdegree{}. At each angle of deposition, various standard
	deviations ($1$\textdegree{}, $2$\textdegree{}, $4$\textdegree{},
	$6$\textdegree{}, $10$\textdegree{}) are used to understand the
	effect of standard deviation on the film growth. In each
	simulation, one million particles are employed to project on to a
	substrate of length one micron. To understand the effect of size of
	particles, simulations are carried out with various particle sizes
	(r = 1.0 \AA{}, 1.5 \AA{}, 2.0 \AA{} and 2.5 \AA{}) at each angle
	of deposition. Standard deviation is kept constant (=
	$1$\textdegree{}) when the effect of size of particles is being
	studied as a function of the angle of deposition. Furthermore,
	since the growth is a random process, ten trials of simulations
	have been carried out for each of these angles of deposition with a
	standard deviation of 1\textdegree{} and disc radius of 1.5 \AA{}.
	The typical height of most of the grown films is about $200$ $nm$.
	The angle of growth, porosity, RMS surface roughness of the
	simulated thin-film deposits have been studied as functions of both
	the angle of deposition and the size of particles. Then the code is
	validated by comparing these results with published theoretical and
	experimental results in Section \ref{Results-2DBD}.

	\subsection{2D Ballistic Deposition on a plane substrate with a
		parallel collimator}
	\label{Method-CGLAD}
\begin{figure}
		\centering
		\includegraphics[width=0.6\linewidth]{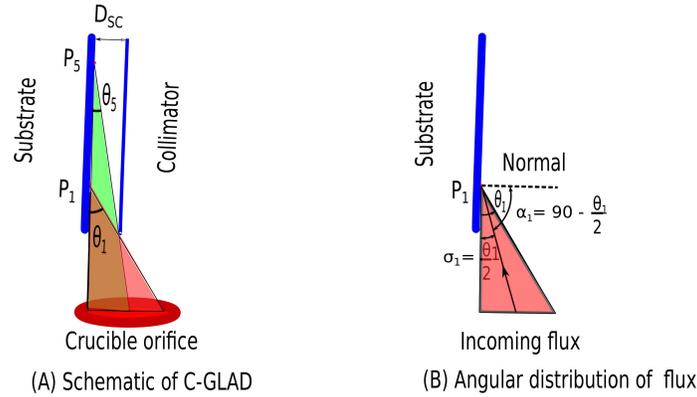}
		\caption{(A) The schematic diagram of a collimated glancing
		angle deposition (C-GLAD) experiment shows the angular
		constrainment of the depositing flux due to a collimator for
		different spots on the substrate. (B) In 	2D-BD simulations,
		particles are deposited at Spot-1 ($P_1$) at an angle chosen by
		sampling from a Gaussian distribution with the given angle of
		deposition $\alpha_1$ and a standard deviation $\sigma_1$.}
		\label{schematic}	
	\end{figure}

	The collimated glancing angle deposition (C-GLAD) technique is
	found to be more effective than a simple GLAD for optimizing the
	microstructure of thin-films
	\cite{garcia2018growth,troncoso2020silver}. In this technique, a
	collimator is placed parallel to the substrate at a suitable
	distance ($D_{SC}$) as shown in Figure \ref{schematic}-(A). With
	this, the angular spread of depositing flux can be constrained and
	the maximum tunability in the morphology of the deposition can be
	achieved. Haque et al. tailored the refractive index of
	silicon-dioxide thin-film using the C-GLAD technique
	\cite{haque2017glancing}. Using this technique, they have also
	shown the tunability of Ag morphology \cite{Divakar}.  They
	demonstrated the change in morphology of Ag deposits from a nearly
	continuous film to nano-islands of varying size with the increase
	in the height of a point on the substrate from its bottom edge.
	
	Five equidistant locations (spots) on the substrate have been
	considered for various characterizations. The distance between any
	two adjacent spots is 10 mm. Figure \ref{schematic}-(A) shows spots
	1  \& 5 ($P_1$ and $P_5$) on the substrate and the remaining spots
	2--4 (or $P_2$, $P_3$, $P_4$) are evenly spaced in between spots 1
	\& 5. It also shows the variation in constrainment of the angular
	spread of the flux reaching these points. The number of atoms
	reaching the substrate decreases substantially with the increase in
	height of the substrate from its bottom edge, due to the increasing
	constrainment on the incoming flux. More details on this collimated
	glancing angle deposition experiment can be found in
	\cite{Divakar}.
	
	Five separate simulations have been performed to grow films with the
	reported thickness mentioned in Table \ref{DivakarParams}. To
	reproduce the features of spot-1, sufficient number of discs 
	with an angular spread of $\theta_1$ have to be deposited. To do this,
	a Gaussian distribution is assumed and deposit sufficient number of
	discs at an angle of deposition $\alpha_1$ (= 90 -
	$\frac{\theta_1}{2}$) with respect to the normal of substrate and
	with a standard deviation of $\frac{\theta_1}{2}$ as shown in
	Figure \ref{schematic}-(B). Similarly, ballistic deposition simulations 
	have been carried out with the angles of deposition
	$\alpha_2$--$\alpha_5$ for the spots 2--5 ($P_2$--$P_5$). The
	parameters used in these simulations are given in Table
	\ref{DivakarParams}. Different morphological features at different
	spots have been observed similar to the C-GLAD experiment. These
	features are discussed in Section \ref{Results-CGLAD}.
	
	\begin{table}[h]
		\centering
		\begin{tabular}{|c|c|c|c|}
			\hline
			Spot ($n$) &	$\alpha_n$ = $90 - \theta_n/2$ & $\sigma_n$ = $\theta_n/2$ &
			Thickness\\ \hline
			1 &	78.2    &   11.8   & 250 nm \\ \hline
			2 &	86.2    &    3.8    & 42 nm \\ \hline
			3 &	87.7    &    2.3    & 26.3 nm \\ \hline
			4 &	88.4    &    1.6    & 18.4 nm \\ \hline
			5 &	88.7    &    1.3    & 16.7 nm \\ \hline
		\end{tabular}
		\caption{Parameters used to reproduce the experimental results
		by Haque et al. \cite{Divakar}. Note that the angle of
		deposition $\alpha_n$ is measured from the normal of the
		substrate and $\sigma_n$ is its standard deviation. Here, $n$
		takes the values from 1--5 for the spots 1--5.}
		\label{DivakarParams}
	\end{table}
	
	\subsection{2D Ballistic Deposition on a sinusoidal patterned
		substrate}
	\label{Method-patternedSub}
	Keller et al. \cite{keller2011polycrystalline} have deposited
	nickel (Ni) on a sinusoidal silicon (Si) substrate to study the
	morphology of thin-film deposition on a sinusoidal patterned
	substrate. They deposited Ni at an angle of (i) 0\textdegree {}
	(ii) 77\textdegree{} with respect to the normal of a sinusoidal
	patterned substrate. The wavelength and the amplitude of the
	sinusoidal substrate are 35 $nm$ and 1.5 $nm$ respectively. They
	noticed that there is high \textit{replicability} or
	\textit{conformity} of the substrate topography on the surface of
	deposited film under normal deposition. This replicability reduces
	with the increase in the height and vanishes at about 120 $nm$. In
	case of glancing angle deposition, they noticed the growth of
	nanorods bunching together on the more exposed areas of
	a sinusoidal substrate. These nanorods further grow into
	anisotropic tilted columns at 7-10 $nm$ height. These tilted 
	columns start merging with the adjacent
	structures at about 10 $nm$. The replicability of the substrate
	topography by the film surface reduces with the increase in height
	and there is no correlation between the substrate and the film
	surface at about 47 $nm$.
	
	2D-BD simulations of thin-film deposition on a sinusoidal substrate
	are carried out to see if the results produced by Keller et al.
	\cite{keller2011polycrystalline} can be reproduced just by
	geometric shadowing effects. In these simulations, circular discs
	with a radius of 1.24 \AA {} are used to represent Ni particles.
	These are initialised uniformly, randomly and then projected 
	towards a substrate as mentioned earlier. Note that
	the experiments used an anisotropic sinusoidal pattern. However,
	they do not mention the amount of anisotropy. For simplicity, this 
	has been considered as an isotropic sinusoidal substrate with the same
	periodicity and amplitude as used by Keller et al. Suitable number
	of particles are projected on to this sinusoidal substrate at (i)
	0\textdegree{} and (ii) 77\textdegree{} to attain the reported
	heights of the film surface. The simulated thin-films are visually
	analysed for the various reported features at various heights.
	These results are presented in Section \ref{Results-patternedSub}.
	
	\section{Results and Discussion}		
	\label{Results}
	\noindent
	2D-BD simulations are carried out to study the morphological
	features of thin-films deposited on (i) a plane substrate, (ii) a
	plane substrate with a parallel collimator and a (iii) sinusoidal
	patterned substrate. The results are presented and compared with
	the available theoretical and experimental results in the following
	Subsections \ref{Results-2DBD}--\ref{Results-patternedSub}.
	
	\subsection{2D Ballistic Deposition on a plane substrate}
	\label{Results-2DBD}
		
		\begin{figure}[h]
			\begin{center}
				\begin{subfigure}{0.5\textwidth}
						\includegraphics[width=\textwidth]{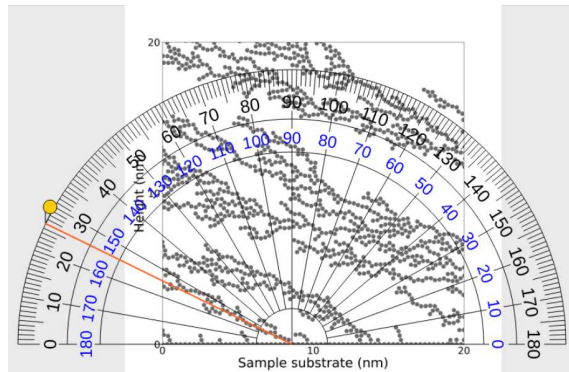}
					\caption{Atomistic view of tilted columns formed at an angle of
						$64$\textdegree{} for an angle of deposition of
						$80$\textdegree{} from the normal. These tree-like structures
						are visible at all the angles of deposition.}
					\label{pattern_angle}
				\end{subfigure}	\newline			
			\begin{subfigure}{0.48\textwidth}		
				\includegraphics[width=\textwidth]{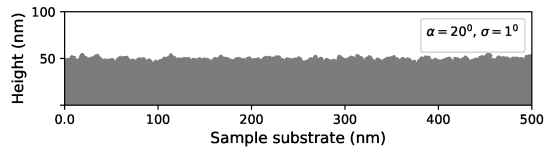} 
				\caption{Continuous thin-film deposited at an angle of 
					$20$\textdegree{} with a standard deviation of 
					$1$\textdegree{} and its angle of growth is $17$\textdegree{}}
				\label{Thinfilm_20deg} 
			\end{subfigure} \hspace{0.35cm}
			\begin{subfigure}{0.48\textwidth}
				\includegraphics[width=\textwidth]{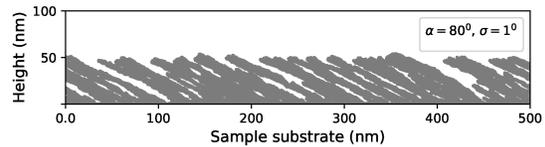} 
				\caption{Porous nano-structured thin-film deposited at 
					$80$\textdegree{} with a standard deviation of 
					$1$\textdegree{} and its angle of growth is 
					$64$\textdegree{}.}
				\label{Thinfilm} 
			\end{subfigure}		
		\caption{(a) A zoomed in figure of tree-like structures of
			connected particles from the simulations are shown. (b) At 
			smaller angles of deposition, frequent branching and lesser 
			spacing between these tree-like structures result in continuous 
			films (c) At glancing angles ($\geq$ 60\textdegree{}), spaces 
			between these patterns increase with increased shadowing effects 
			and result in tilted columns.}
		\label{}	
		\end{center}
	\end{figure}		
	
	The ``tree-like structures'' reported earlier at the atomic scales 
	\cite{dirks1977columnar, meakin1988invited,	mansour2019ballistic} 
	have been observed in all the simulations for all the angles
	deposition carried out in the present study as seen in
	Figure.\ref{pattern_angle}. The height of these structures
	increases with deposition. The branching of the tree-like
	structures is more and they are more closely packed at smaller
	angles of deposition ($20$\textdegree{}, $30$\textdegree{}).
	Therefore for these angles, a thin-film looks continuous as shown
	in Figure \ref{Thinfilm_20deg}. The shadowing effects in a
	ballistic deposition, increase with the increase in the angle of
	deposition. With the increasing shadowing effects, the space
	between the `tree-like structures increases. One can see these
	structures as closely spaced, thin, tilted columns for the angles
	of deposition $40$\textdegree{}, $50$\textdegree{}. The size and
	space between these tilted columns increase with the angle of
	deposition and also with the standard deviation. The tilted columns
	are easily discernable for angles of deposition $ \geq
	60$\textdegree{}. Figure \ref{Thinfilm} shows the tilted columns
	formed at an angle of deposition $80$\textdegree{} with a standard
	deviation of $1$\textdegree{}. The angle of growth of these
	tree-like structures or tilted columns is measured using an online
	protractor tool \cite{onlineprotractor}. In this subsection, angle
	of growth, porosity and RMS surface roughness of the simulated
	thin-films on a plane substrate are discussed as a functions of the
	angle of growth and size of the particles. These are compared with
	published results to validate the code.
	
	\subsubsection{Angle of growth of thin-film} 
	\label{AngleGrowth} 
	With simple geometric effects, the earlier two-dimensional
	ballistic deposition simulations  could reproduce several key
	observations such as the angle of growth ($\beta$) is less than the
	angle of deposition ($\alpha$) and a monotonic decrease in the
	density of thin-film with the increase in the angle of deposition
	\cite{henderson1974simulation, dirks1977columnar,
	meakin1988ballistic}. Various rules have been proposed to describe
	the relationship between $\alpha$ and $\beta$. We compare our
	results with widely used ``tangent rule'' 
	($tan \alpha = 2 \hspace{1mm} tan \beta$) \cite{tangentrule},
	``cosine rule'' 
	($\beta = \alpha - arcsin \left(\frac{1-cos\alpha}{2} \right)$)
	\cite{cosRule}.
	
	\begin{figure}
		\centering
		\includegraphics[width=0.7\linewidth]{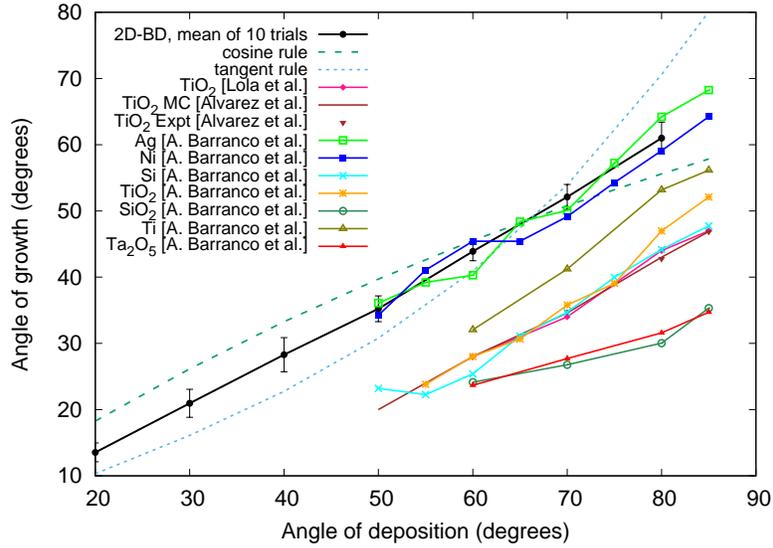}
		\caption{The angle of growth increases with the angle of
		deposition. The mean values of angle of growth are estimated
		for ten trials at each angle of deposition from
		$20$\textdegree{}--$80$\textdegree{} with a standard deviation
		of $1$\textdegree{}.}
		\label{variance}
	\end{figure} 
	
	The tilted  columns of a thin-film are mostly parallel to each
	other. The angle of  growth of these tilted columns is considered
	to be the angle of growth  of the thin-film.  Few columns may not
	be parallel to the remaining due to ``column extinction'' at some
	points due to shadowing effects. The height and orientation of any
	specific column depend on the orientation and height of the
	adjacent shadowing columns. In the present study, this issue is
	taken care of by considering an average value of angles of many
	tilted columns for each simulation. Further, at each angle of
	deposition with a standard deviation of $1$\textdegree{}, ten
	simulations with different random number seeds are carried out. The
	angle of growth from these ten simulations are obtained as
	described before and their average value is considered. These
	angles of growth and their standard deviations are estimated and
	compared with the available theoretical and experimental results in
	Figure \ref{variance}. The angle of growth increases with the angle
	of deposition. The values of angle of growth are also estimated at
	each angle of deposition with standard deviations of
	$1$\textdegree{}, $2$\textdegree{}, $4$\textdegree{},
	$6$\textdegree{}, $10$\textdegree{} to understand the effect of
	standard deviation in the angle of deposition. It is found that
	there is no significant change in the angle of growth with the
	change in standard deviation at a particular angle of deposition.
	
	\begin{figure}[t]
		\begin{tabular}{c c}
			\includegraphics[width=0.48\linewidth]{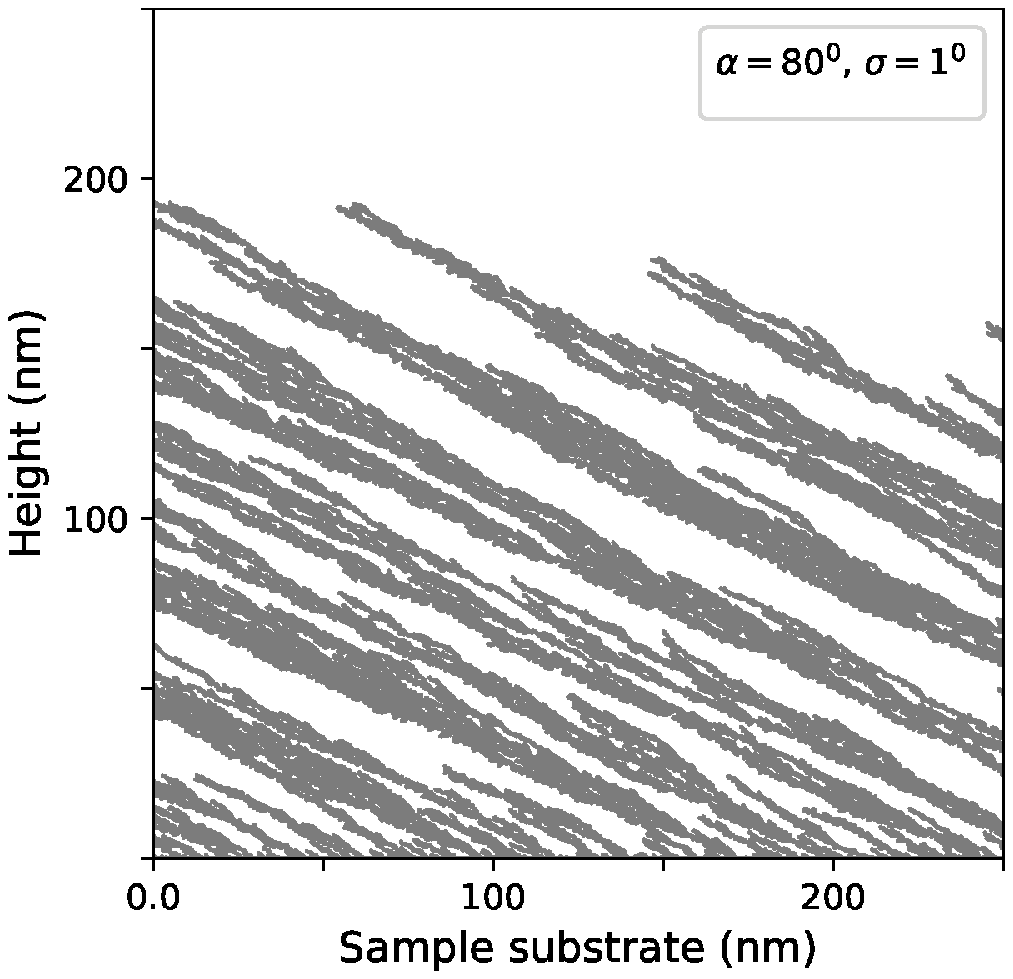} &
			\includegraphics[width=0.48\linewidth]{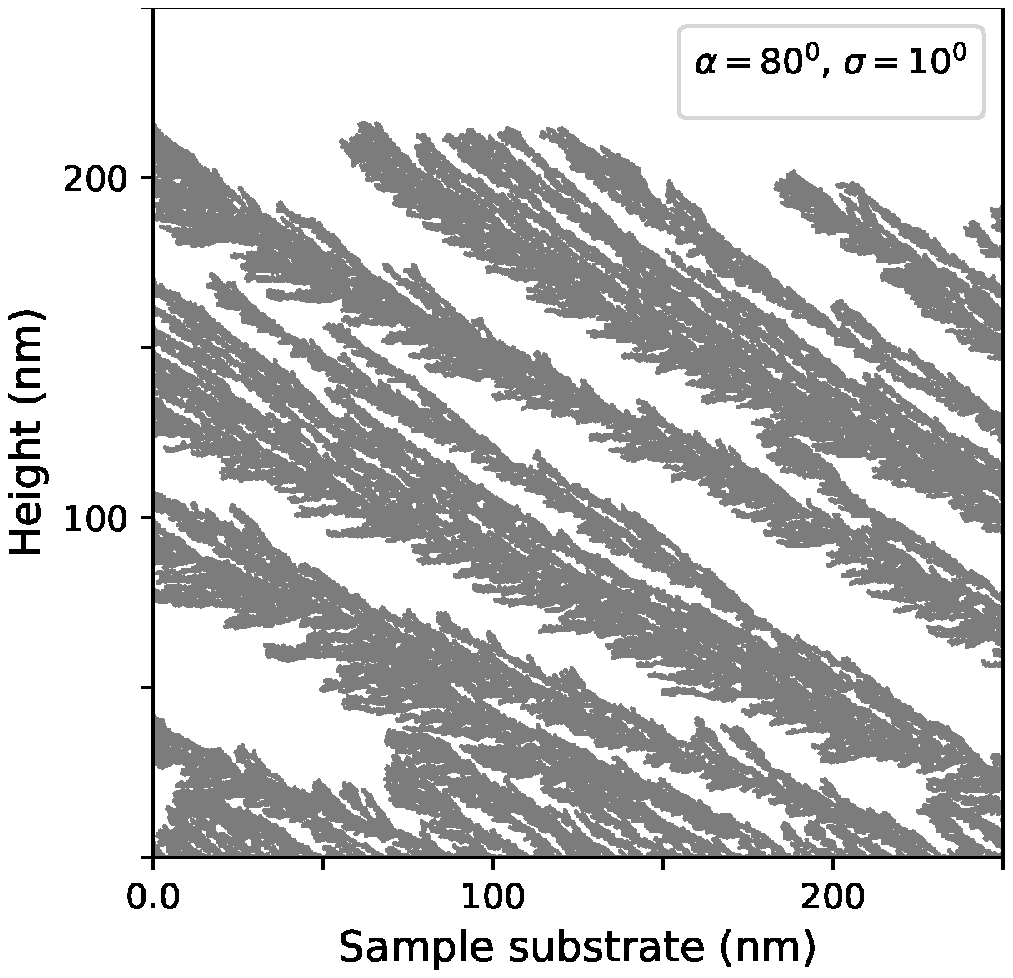}
		\end{tabular}	
		\caption{For a particular angle of deposition, angle of growth
		remains constant with the increase in standard deviation, but
		the width of tilted columns increases.}
		\label{columns}
	\end{figure}
	
	Figure \ref{variance} shows the relationship between the angle of
	deposition and the angle of growth obtained from the simulations as
	described above and compares it with the empirical rules
	\cite{tangentrule,cosRule} and experiments \cite{material1,
	material2, Lola, review2016}. The mean values of angles of growth
	from the simulations lie in between the values estimated by tangent
	and cosine rules. They are found to match closely with the
	experimental results \cite{material2,review2016} for pure metals
	having a cubic lattice structure. They do not match for others
	either due to preferential direction in bonding or experimental
	conditions. The angle of growth values from our simulations match
	the 2D and 3D simulations by Meakin et al
	\cite{meakin1988ballistic} within a range of $1$--$3$\textdegree{}.
	We do not observe any significant change in the angle of growth with
	the change in the standard deviation, but there are morphological
	changes like the shape of thin-film deposition, the thickness of
	tilted columns. Figure \ref{columns} shows the increase in the
	thickness of tilted columns with an increase in the standard
	deviation.
	
	\subsubsection{Porosity} \label{Porosity}
	Porous thin-films have a broad spectrum of applications due to
	their adequate characteristics such as high resistance to thermal
	shock, low thermal conductivity, etc. Using oblique angle
	deposition (OAD), one can engineer the required porosity for
	various applications. In two-dimension:
	
	\begin{equation}
	Porosity = \frac{Void{\hspace{1.5mm}}area}{Total{\hspace{1.5mm}} area} 
	= \frac{Total{\hspace{1.5mm}}area - 
		Area{\hspace{1.5mm}}occupied{\hspace{1.5mm}}by 
		{\hspace{1.5mm}}discs}{Total{\hspace{1.5mm}} area}
	\end{equation}
	
	\begin{figure}
		\centering
		\includegraphics[width=0.9\linewidth]{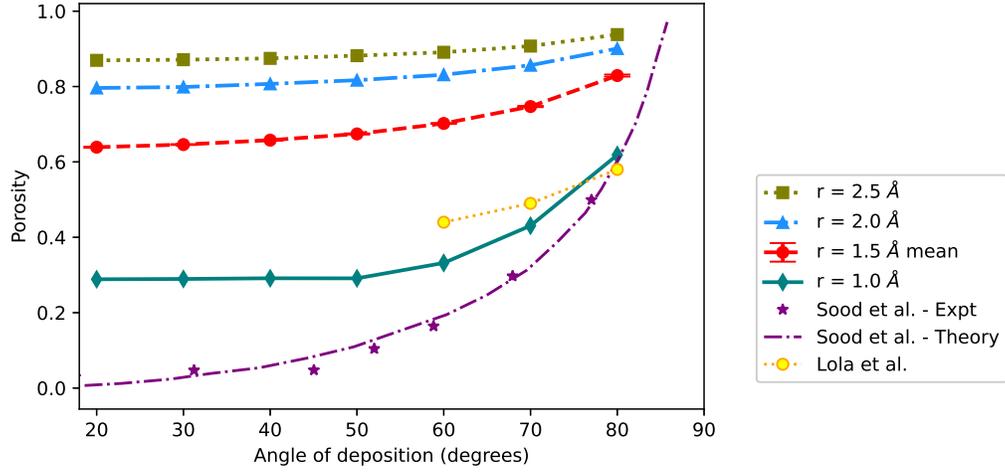}
		\caption{Porosity increases with the angle of deposition and also 
			with the radius (r) of the particles.}
		\label{porosity_fig}	
	\end{figure}
	
	\noindent
	Porosity (P) increases with the angle of deposition
	\cite{Alvarez,Lola,sood,MD2013} and remains constant with the
	height of the film \cite{porDistr2020}. Figure \ref{porosity_fig}
	shows the porosity (P) as a function of the angle of deposition
	($\alpha$) and confirms that porosity increases with the increase
	in the angle of deposition. It can also be seen  that porosity
	increases with an increase in the radius (r) of the particle used
	in the simulation, which is a geometric effect.  The effect of
	standard deviation ($\sigma$) is found to be insignificant at any
	specific angle of deposition ($\alpha$). Earlier, Sood et al. 
	have used the following expression (\ref{sood_formula}) with a 
	``fitting parameter'' ($c$) to fit
	their experimental results with their Monte Carlo simulations for
	indium tin oxide (ITO) \cite{sood}.
	
	\begin{equation}\label{sood_formula}
	P = \frac{\theta tan \theta}{c + \theta tan \theta} \hspace{1cm} 
	\textit{Where c = 8.32 and $\theta$ is the angle of deposition.}
	\end{equation}	
	
	The results for porosity,
	reported in the present study, follow the trend observed in
	experiments. In most of the earlier theoretical studies of
	ballistic deposition, scaling behaviour of porosity is studied
	using the particles with unit size \cite{kim1977computer,
	meakin1987restructuring, krug1989microstructure, tait1990ballistic,
	banerjee2014surface, mal2016surface}. To the best of our knowledge
	the effect of size of particles on the porosity has not been
	reported. Here, it is suggested that in simulations, one needs to
	use the discs of size comparable to that of particles of practical
	interest, for a quantitative match with the experimental results.
	
	\subsubsection{RMS Surface Roughness} 
	\label{Roughness}
	Roughness is defined as how the height ($h(r)$) of a surface
	deviates from its average height ($\bar{h}$). It is measured by
	finding the root mean square (RMS) variation of the height and is
	expressed \cite{luo} as:
	\begin{equation}
	R_q = \sqrt{\frac{\sum_{i=1}^{N}(h_i - \bar{h})^2}{N}}
	\end{equation}
	
	\begin{figure}
		\begin{center}
			\begin{subfigure}{0.49\linewidth}
			\includegraphics[width=\linewidth]{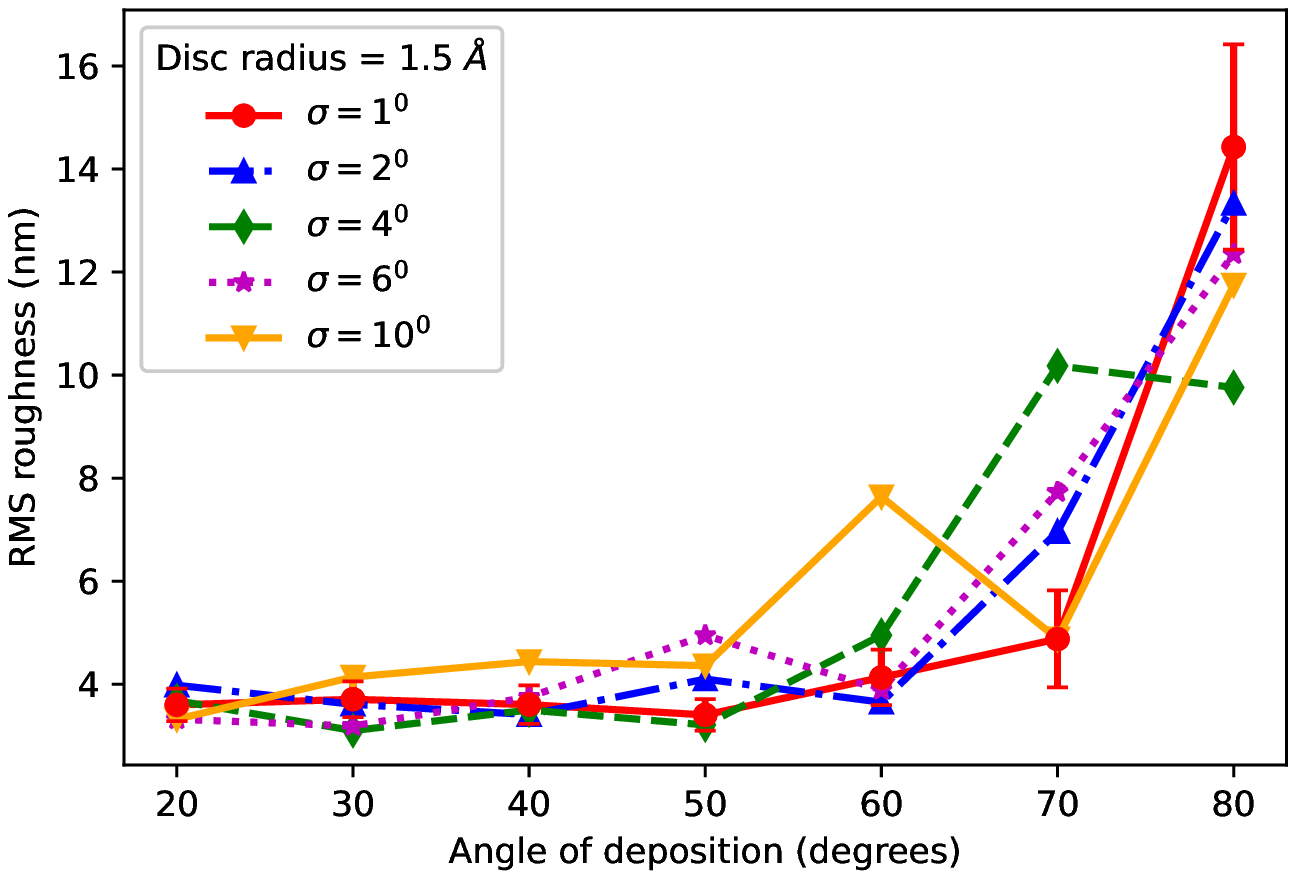}
			\caption{}
				\label{roughness_variance}	
			\end{subfigure}	\hspace{1mm}		
			\begin{subfigure}{0.49\linewidth}			
				\includegraphics[width=\linewidth]{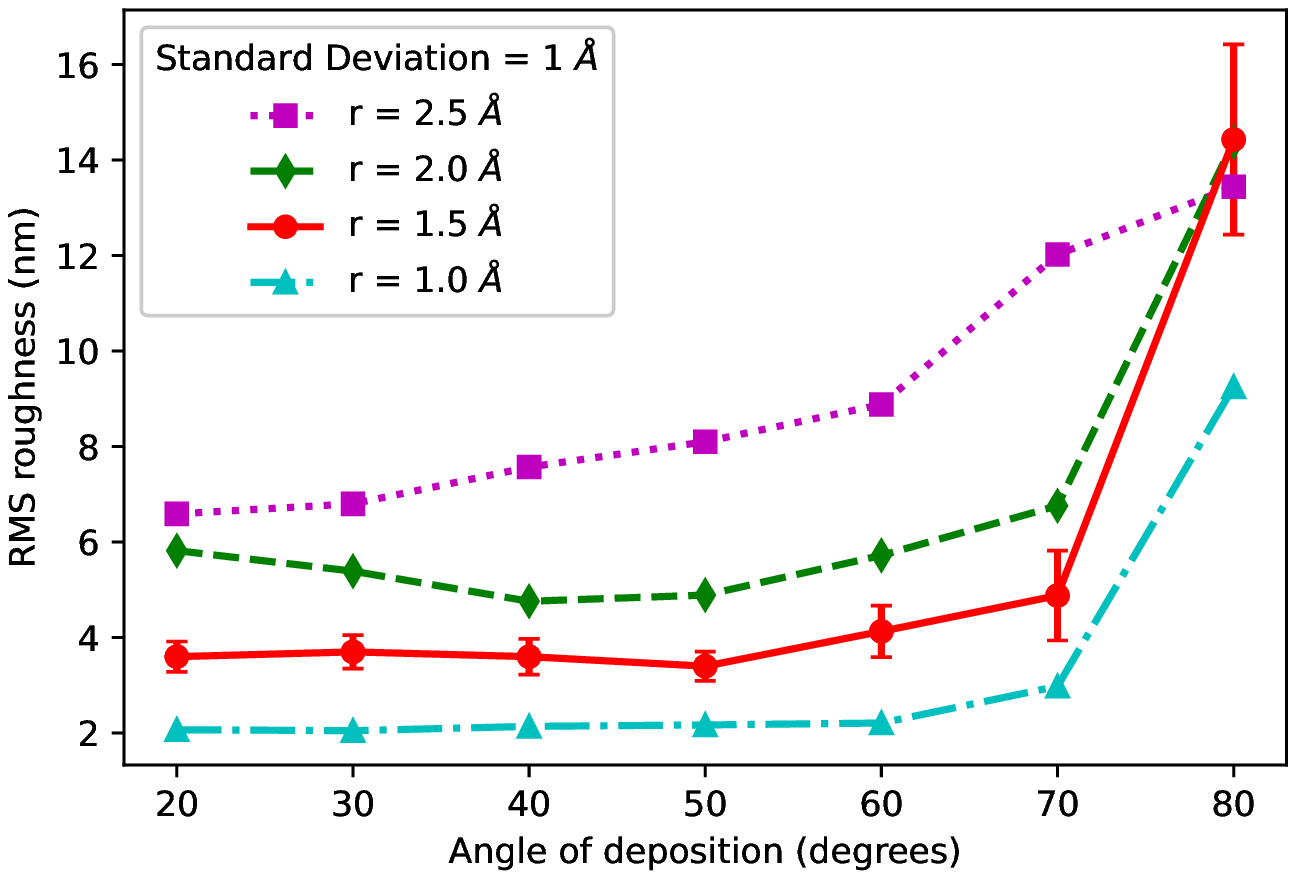}
				\caption{}
				\label{roughness_sizeComp}	
			\end{subfigure}	
		\caption{(a) The RMS surface roughness increases with the increase in
			the	angle of deposition and the trend does not depend on the
			standard deviation in the angle of deposition. (b) It increases with the increase in size of particles.}		
		\end{center}	
	\end{figure}
	
	\noindent
	Here, mean height ($\bar{h}$) is estimated for the sampling length
	of the substrate. To measure the RMS surface roughness ($R_q$), the
	surface in the range of sampling length is divided into small bins
	whose size is equal to the size of the discs. Then mean height
	($\bar {h}$) and the root mean square deviation of all heights (RMS
	surface roughness) are measured for all the simulations carried out
	in the present study.	
	
	Figure \ref{roughness_variance} shows the RMS surface roughness as
	a function of the angle of deposition. In general, It is observed
	that RMS surface roughness increases with the increase in the angle
	of deposition. To understand the effect of standard deviation on
	RMS surface roughness, simulations are carried out with discs of a
	fixed radius equal to 1.5\AA{} using standard deviations
	1\textdegree{}, 2\textdegree{}, 4\textdegree{}, 6\textdegree{} and
	10\textdegree{}. It can be seen that there is no specific trend for
	the change in RMS surface roughness with the change in the standard
	deviation in the angle of deposition. To estimate the statistical
	error, ten simulations are carried out with $\sigma =
	1$\textdegree{} and the average values are plotted with their
	statistical error. The change in RMS surface roughness is
	insignificant for $\alpha <60$ \AA{}. Figure
	\ref{roughness_variance} shows that the values of RMS surface
	roughness range from 2-16 nm closely match with the reported range
	2-14 nm from the surface analysis of OAD evaporated thin-films by
	atomic force microscopy (AFM) \cite{review2016, luo,
	bouaouina2018nanocolumnar}.
	
	To understand the effect of size of the particles on RMS surface
	roughness, simulations are carried out with the discs of radii
	equal to 1.0\AA{}, 1.5\AA{}, 2.0\AA{} and 2.5\AA{}. The standard
	deviation (=1\AA) is kept constant at each angle of deposition. 
	Figure \ref{roughness_sizeComp} shows that the RMS surface
	roughness increases with the increase in the angle of deposition
	for a given size of the particles. It can also be seen that the RMS
	surface roughness increases with the increase in the size of 
	particles. Here, the mean values of RMS surface roughness with
	their statistical error are plotted for the particles of radius
	equal to 1.5\AA{}.
	
	\subsection{2D Ballistic Deposition on a plane substrate with a
		parallel collimator}
	\label{Results-CGLAD}
	
	\begin{figure}
		\begin{center}
			\begin{subfigure}{0.5\textwidth}
				\centering
				\includegraphics[width=\textwidth]{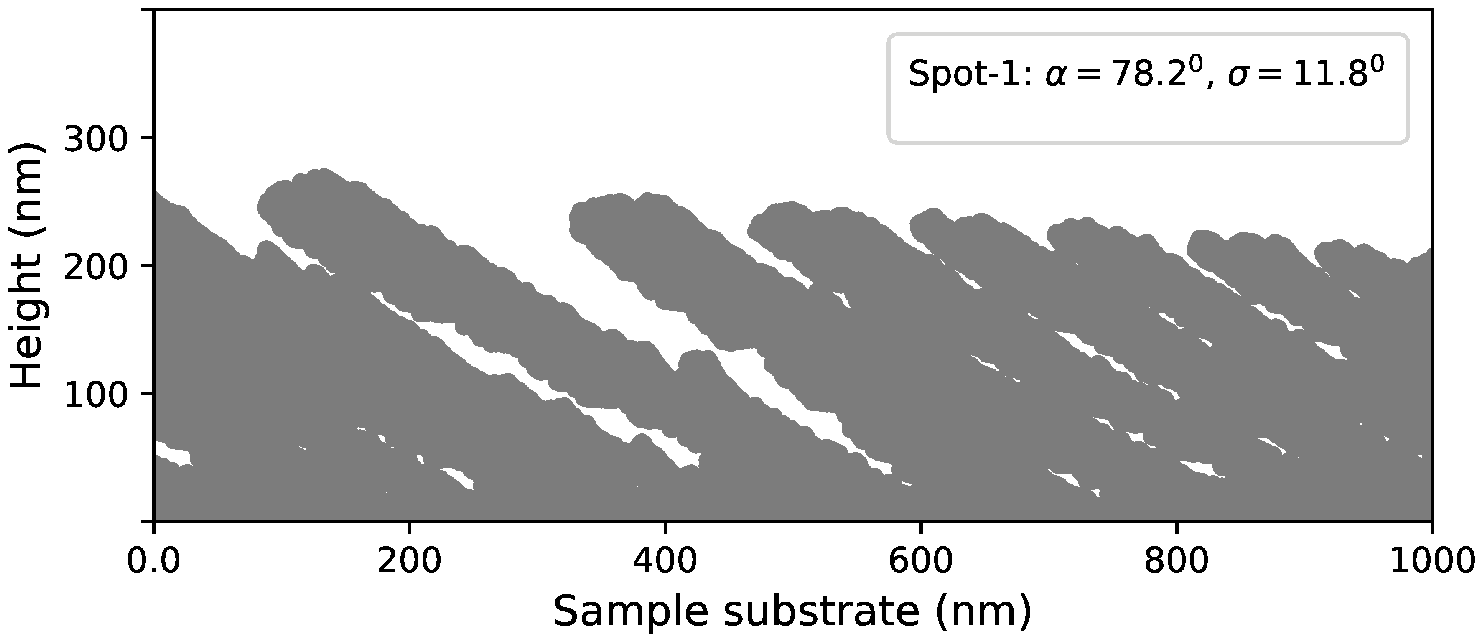}
			\end{subfigure}
		\end{center}
		\begin{subfigure}{0.48\linewidth}
			\includegraphics[width=\textwidth]{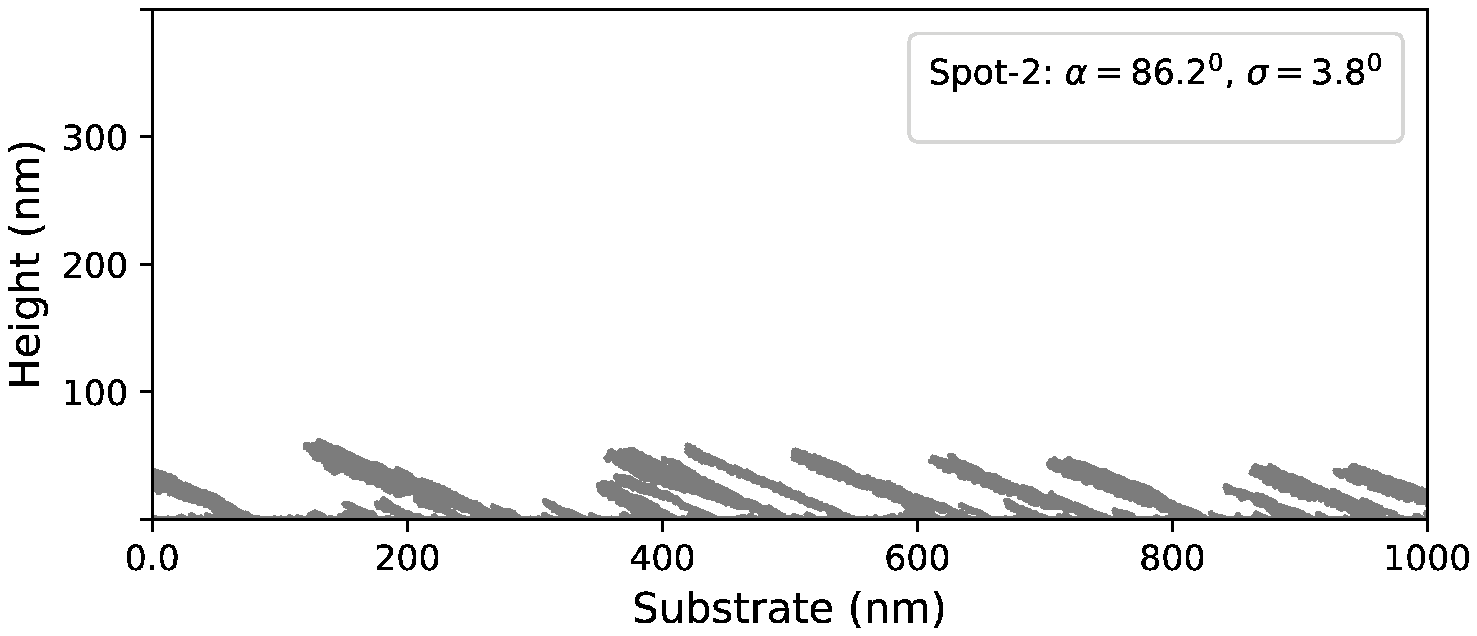}
		\end{subfigure}
		\begin{subfigure}{0.48\linewidth}
			\includegraphics[width=\textwidth]{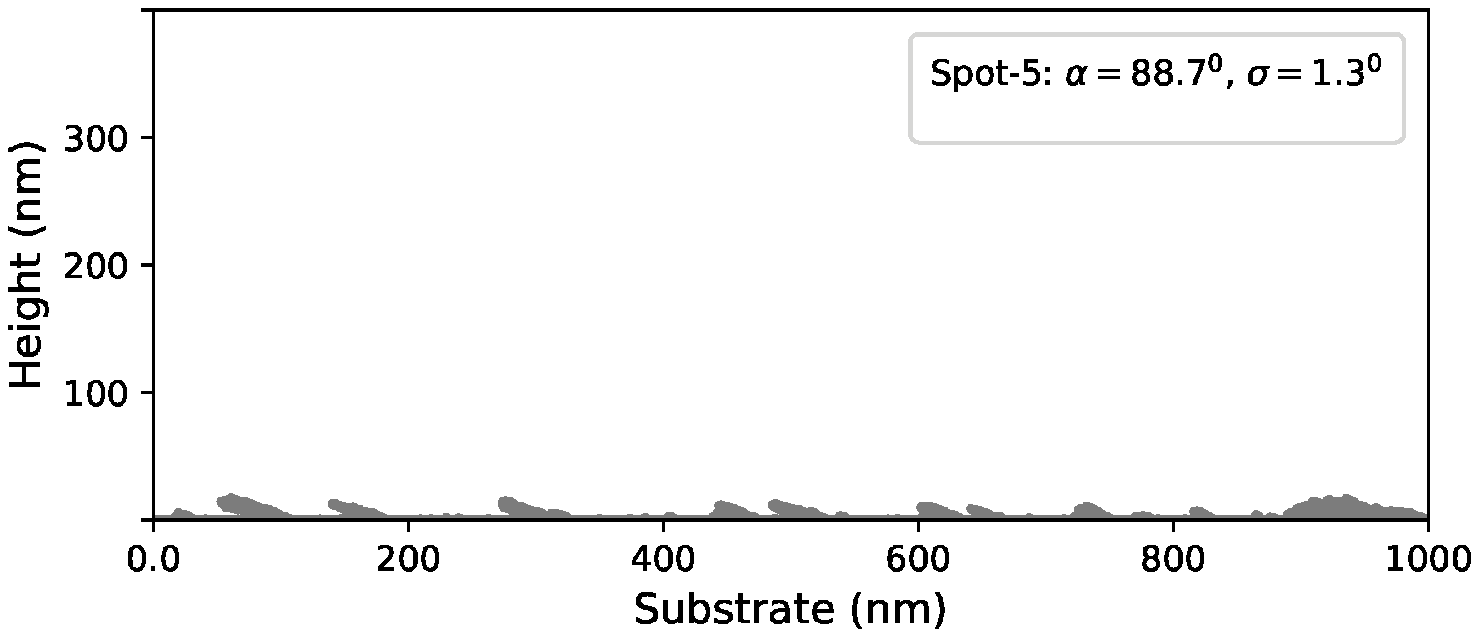}
		\end{subfigure}
		\caption{Continuous thin-film with protrusions, elongated
			nano-rods and isolated nano-islands produced by 2D-BD
			simulations for the spots 1, 2 and 5 of C-GLAD experiment.}
		\label{DivakarExpt}
	\end{figure}
	
	The 2D-BD simulations reproduce similar structures as obtained by
	Haque et al. \cite{Divakar} and can be seen in Figure
	\ref{DivakarExpt}. In the initial phase of the thin-film growth,
	small stable clusters of atoms grow into three-dimensional
	nano-islands. Later, these nano-islands act as growth seeds of
	nano-rods. These nano-rods further grow to tilted columns resulting
	in continuous thin-films with protrusions. At spot-1, the film is
	almost continuous due to larger deposition. At spot-2, it has
	nanorod structures with reduced deposition. From spot-3 onwards
	with further reduction in the beam deposition, only discrete
	nano-islands form and they do not grow large enough to merge and
	form a continuous film within the allowed deposition time. The
	observed sizes are compared with the reported sizes 	from the
	C-GLAD experiment in Table \ref{CGLD-2DBD comp}. Porosity for the
	spots- 1 \& 2 are estimated to be 0.79 and 0.96. RMS surface
	roughness of these first and second spots are 11.25 nm, 12.04 nm
	respectively for the sampling lengths from 0.0--200 $nm$. Porosity
	and RMS surface roughness characterizations are not considered for
	the other spots as they are very small individual nano-islands. The
	number of the individual nano-islands per micron length for the
	spots 3--5 are about 12, 16, 13  and 15, 21, 18 in the 2D-BD
	simulations and C-GLAD respectively.
	
	\begin{table}[h!]
		\centering
		\begin{tabular}{|c|c|c|c|c|c|}
			\hline
			\multirow{3}{*}{Spot} & Nature & Average & Average & Average spaces & Average
			spaces \\
			& of the & Reported size & Observed size & between C-GLAD & between 2D-BD \\
			& nanostructure & in C-GLAD & in 2D-BD  & nanostructures & nanostructures\\
			\hline
			\multirow{2}{*}{1} & Continuous & \multirow{2}{*}{--} &  \multirow{2}{*}{--} &
			protrusions are & protrusions are \\
			& with protrusions & & & contiguous & contiguous \\ \hline
			\multirow{2}{*}{2} & elongated  & \multirow{2}{*}{$135 \pm 15$ nm} &
			\multirow{2}{*}{116 nm}  & \multirow{2}{*}{$53 \pm 19$}   & \multirow{2}{*}{$83
				\pm 16$ nm} \\ 
			&nanorods &&&& \\ \hline
			\multirow{2}{*}{3} & isolated  &\multirow{2}{*}{$83 \pm 10$ nm} &
			\multirow{2}{*}{55 nm}  & \multirow{2}{*}{$44 \pm 15$} &  \multirow{2}{*}{$48
				\pm 15$ nm} \\ 
			&nano-islands&&&& \\ \hline
			\multirow{2}{*}{4} & isolated  & \multirow{2}{*}{$40 \pm 8$ nm} &
			\multirow{2}{*}{30 nm}  &\multirow{2}{*}{$33 \pm 5$} & \multirow{2}{*}{$51 \pm
				27$ nm} \\ 
			&nano-islands&&&& \\ \hline
			\multirow{2}{*}{5} & isolated  & \multirow{2}{*}{$26 \pm 4$ nm} &
			\multirow{2}{*}{30 nm} & \multirow{2}{*}{$35 \pm 7$} & \multirow{2}{*}{$56 \pm
				19$ nm} \\
			&nano-islands&&&& \\ \hline
		\end{tabular}
		\caption{Comparison of sizes of nanostructures reported by Haque et
			al. \cite{Divakar} and observed in the present two-dimensional
			ballistic deposition (2D-BD) simulations.}
		\label{CGLD-2DBD comp}
	\end{table}
	
	The 2D-BD simulations could reproduce the morphology of five
	equidistant spots in the C-GLAD experiment ranging from continuous
	thin-film to small individual nano-islands. There is a decent
	quantitative match for the average sizes of the nano-structures and
	the average spaces between them. The values of the linear density
	of number of nano-islands obtained in the 2D-BD simulations for
	spots 3--5 are within the range of their values obtained in the
	C-GLAD experiment. The results of 2D-BD simulations have a good match 
	with the results of C-GLAD experiment.
	
	\subsection{2D Ballistic Deposition on a sinusoidal patterned substrate}
	\label{Results-patternedSub}
	
	\begin{figure}[h]
		\begin{center}			
		\begin{subfigure}{0.48\linewidth}
		\includegraphics[width=\textwidth]{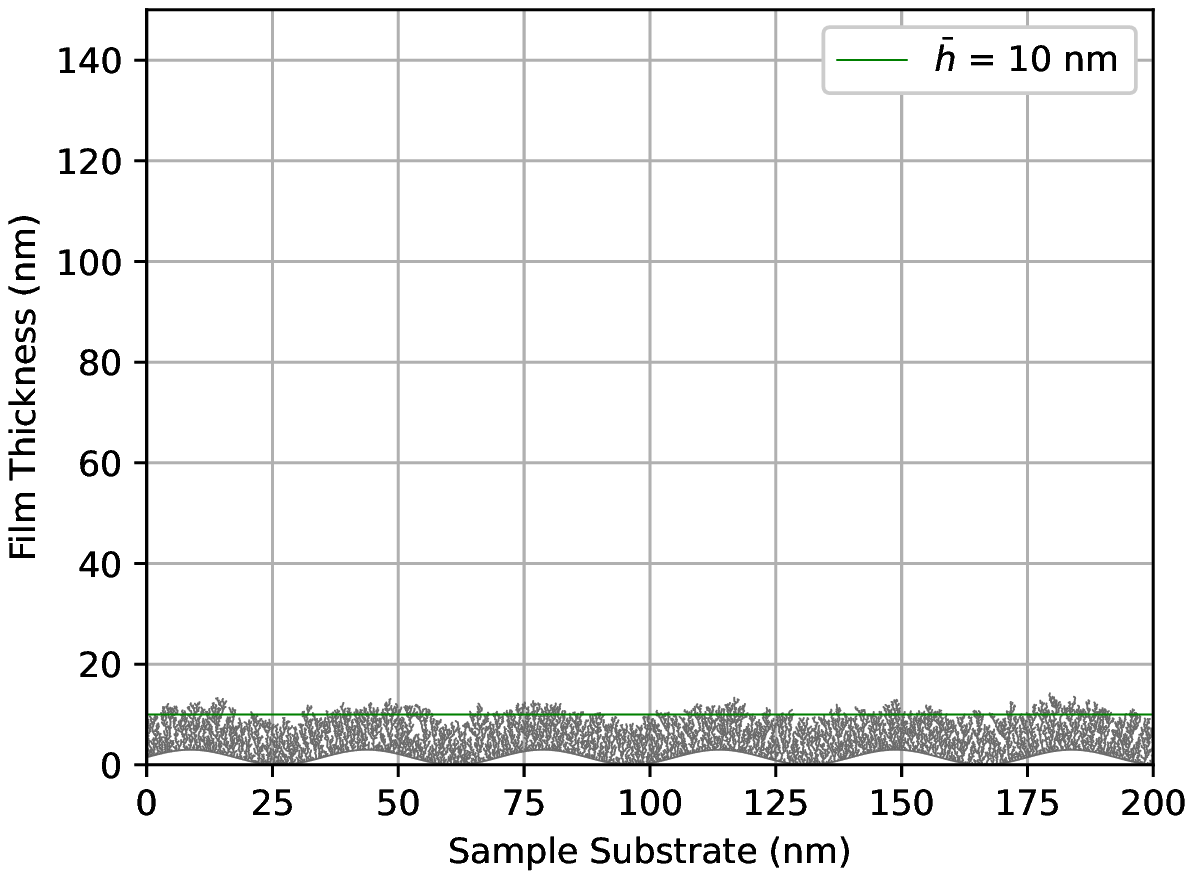}
		\caption{}
		\label{10nm_normDep}
		\end{subfigure}
		\begin{subfigure}{0.48\linewidth}
		
		\includegraphics[width=\textwidth]{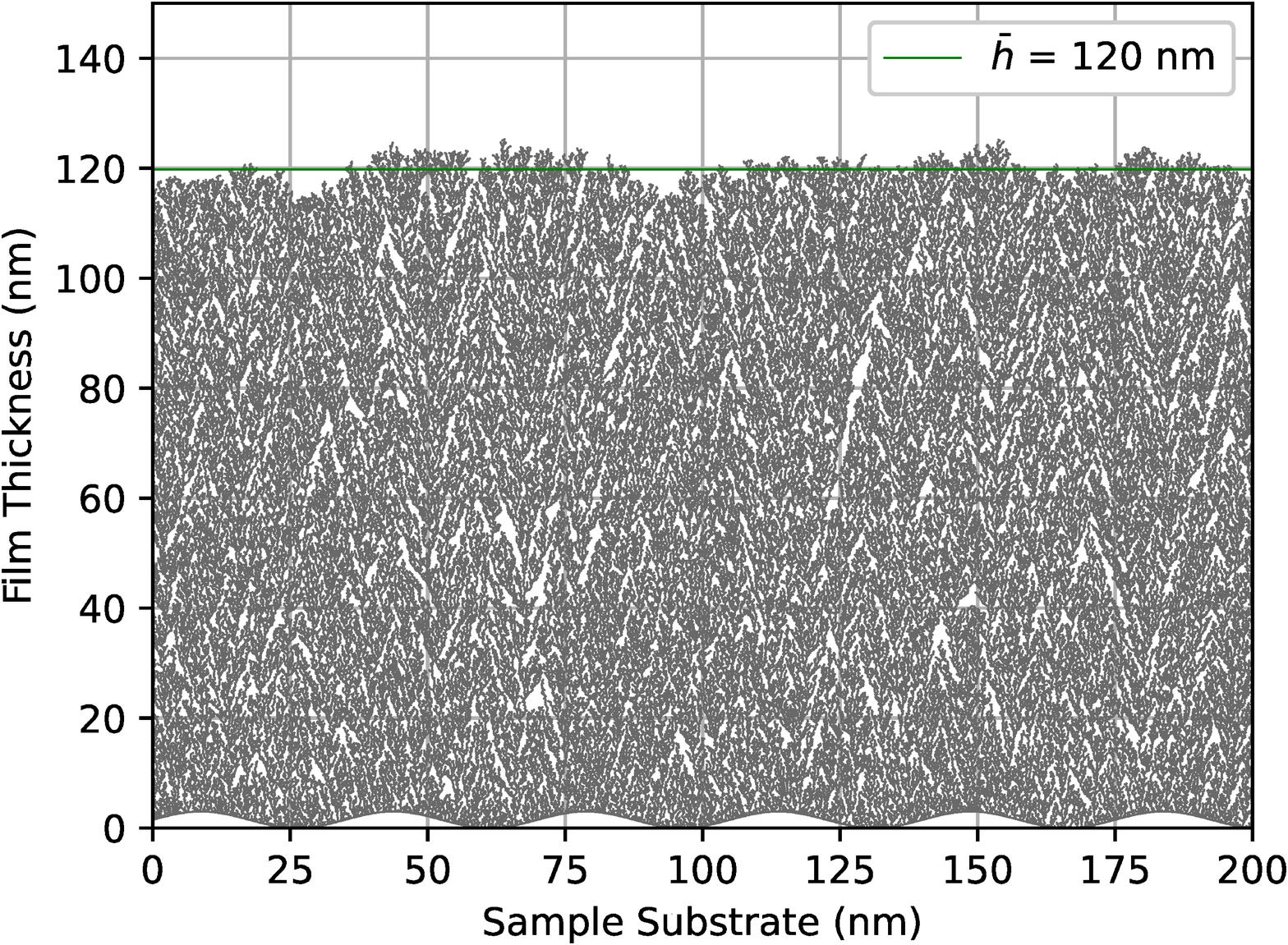}
		\caption{}
		\label{120nm_normDep}
		\end{subfigure}
		\begin{subfigure}{0.48\linewidth}
		\includegraphics[width=\textwidth]{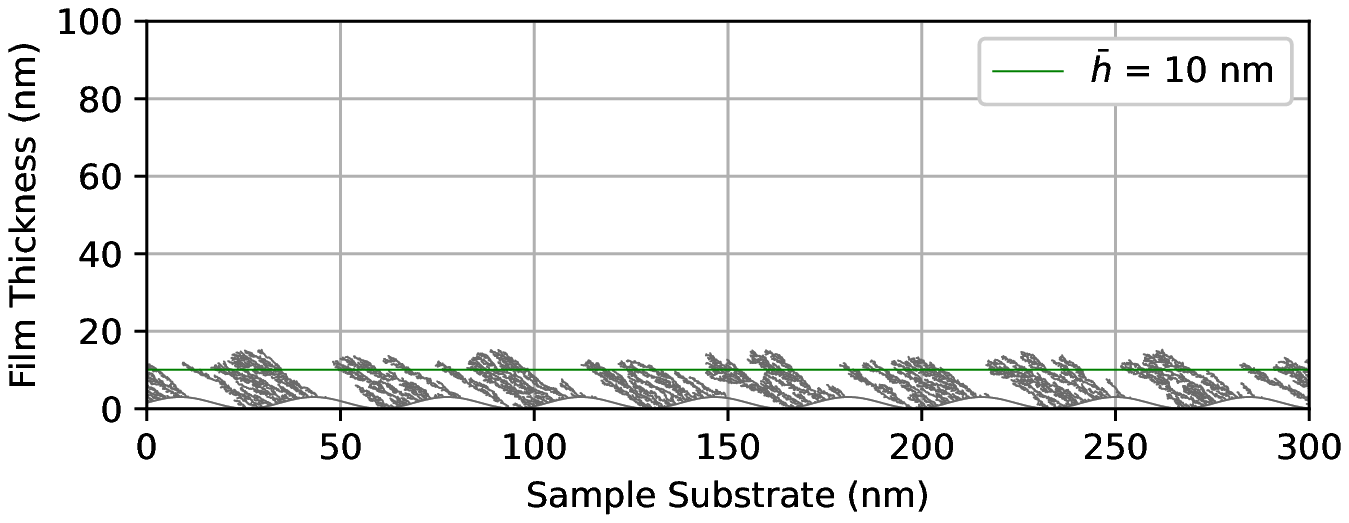}
		\caption{}
		\label{10nm_OAD}
		\end{subfigure}
		\begin{subfigure}{0.48\linewidth}
		\includegraphics[width=\textwidth]{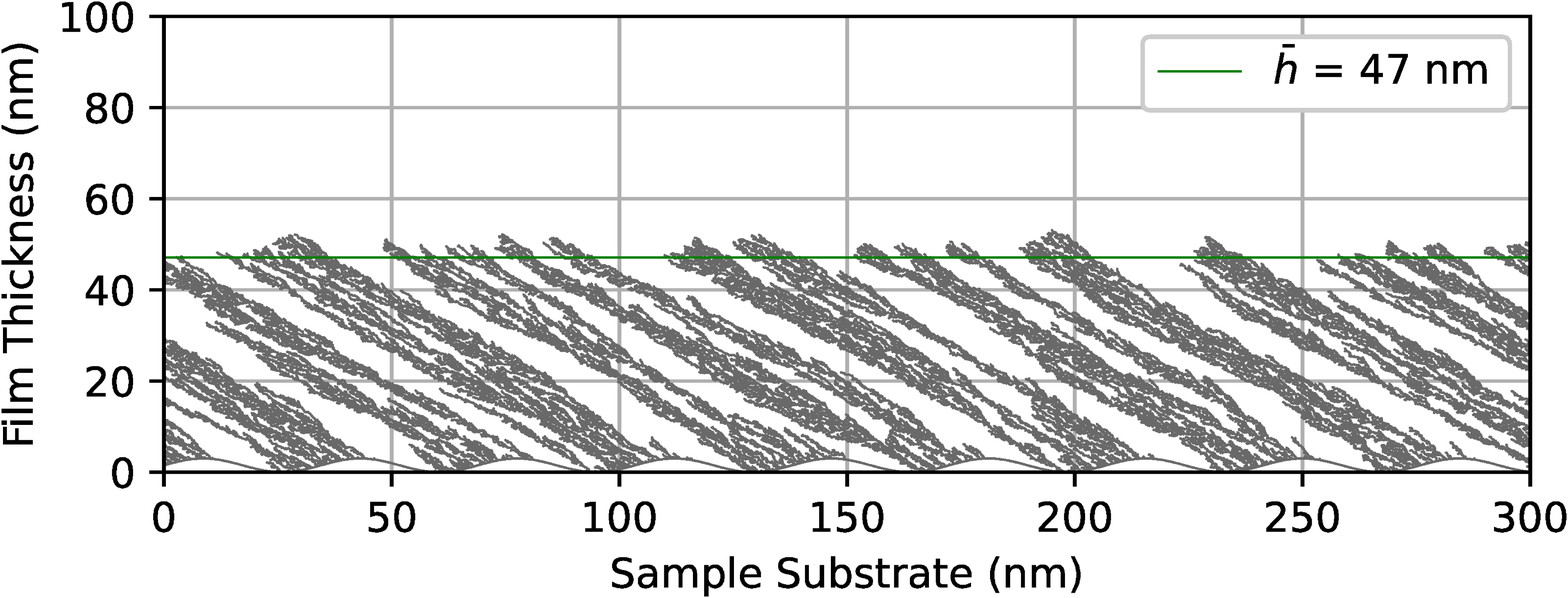}
		\caption{}
		\label{47nm_OAD}
		\end{subfigure}
		\caption{(a \& b) Replicability of a sinusoidal patterned
			substrate topography by the surface of a thin-film gradually
			decreases with the increase in height and vanishes above 120
			$nm$ under normal deposition. (c \& d) In OAD, this
			replicability rapidly decreases with the increase in height
			and	vanishes at about 47 $nm$. The reported preferential
			growth mode of depositing flux on to the more exposed areas
			of sinusoidal substrate can be seen in OAD.}
		\label{pattern_dep}		
	\end{center}
	\end{figure}
	
	2D-BD simulations of thin-film deposition are carried out at
	0\textdegree{}, 77\textdegree{} on a sinusoidal substrate. In case
	of normal deposition, there is high replicability or conformity of
	substrate topography by the film surface and it is observed up to a
	height of about 10 $nm$ as shown in Figure \ref{pattern_dep}(a).
	The reported reduction in both the replicability and amplitude of 
	the surface modulation with the increase in height of the film 
	surface have been observed. This is due to the increase in the 
	surface roughness with increasing height. The replicability gradually
	decreases up to a height of 120 $nm$ as shown in Figure
	\ref{pattern_dep}-(b) and it vanishes at further heights. Here,
	$\bar{h}$ is the mean height of the film surface.
	
	In the case of oblique angle deposition (OAD) at 77\textdegree{},
	the reported preferential growth of nanorods on the more exposed
	areas of ripples that are facing the depositing flux is
	observed. This preferential growth is shown in Figure
	\ref{pattern_dep}-(c). The nanorods further grow to form
	anisotropic tilted columns. When the height of thin-film reaches
	about 7--10 $nm$, they start growing over the adjacent peaks of the
	substrate. This results in the closure of gaps between the more
	exposed faces of the rippled substrate. With this, the
	replicability reduces rapidly and vanishes at about 47 $nm$ as
	shown in Figure \ref{pattern_dep}-(d). Keller et al.
	\cite{keller2011polycrystalline} predicted the angle of growth from
	tangent rule \cite{tangentrule} to be 71\textdegree{} for an angle
	of deposition 77\textdegree{}. In their experiment, they found a
	lesser angle of growth and is 50\textdegree{}. The angle of growth
	obtained in the present simulations is $56 \pm 5$\textdegree{},
	which is a better match with the experimental result. This shows
	that 2D-BD simulations can predict the angle of growth better than
	the conventional empirical rules. The reported long-lasting growth
	mode on the more exposed faces of the ripples is also observed in the
	present simulations.
	
	\section{Conclusion} \label{Conclusions}
	A two-dimensional ballistic deposition code, ``2D-BD'' based on
	geometric shadowing effect has been	developed to to study the
	growth of porous nano-structured thin-films. A criterion for the
	applicability of 2D-BD simulations has been developed. The code is
	validated by comparing with the published results for a plane
	substrate and then it is further validated by applying it to two
	experiments carried out using a plane substrate with a collimator
	and a sinusoidal patterned substrate. For plane substrate,
	morphological features such as the angle of growth ($\beta$),
	porosity (P) and surface roughness ($R_q$) are studied as functions
	of the angle of deposition ($\alpha$) and the size of the
	particles. It is found that $\beta$, P, $R_q$ increase with the
	increase in the angle of deposition as reported by the earlier
	studies. The angle of growth ($\beta$) remains constant with the
	change in size, whereas the porosity and surface roughness increase
	with the increase in the size of the particle.
	
	For a substrate with a parallel collimator, various morphological
	features of the deposited films ranging from continuous thin-film
	to small individual nano-islands are reproduced. The average sizes
	of nanostructures and the average spacing between them from 2D-BD
	simulations show a good match with the results obtained in the
	C-GLAD experiment. The linear density of number of nano-islands
	obtained in 2D-BD simulations for the spots 3--5 are within the
	range of the values obtained in the C-GLAD experiment. For a
	sinusoidal patterned substrate, the reported replicability of the
	substrate topography by the surface of the thin-film is reproduced
	up to a height of 120 $nm$ under normal deposition. In case of
	oblique angle deposition with the sinusoidal substrate, this
	replicability reduces rapidly and vanishes by a height of 47 $nm$.
	The experimentally reported preferential and long-lasting growth
	mode of depositing flux on to the more exposed areas of a rippled
	substrate is ascertained.The 2D-BD simulations can predict the 
	angle of growth of deposits on patterned substrate better than 
	the conventionally used empirical rules. 
	
	2D-BD simulations, though not as exact as MD simulations, are 
	relatively fast and can handle larger sizes of thin-film deposits. 
	They can be used to study the details of experimentally observable 
	morphological features such as the angle of growth, porosity, RMS 
	surface roughness on both the plane and patterned substrates for 
	the thin-films of heights of a few hundred nanometres deposited 
	over a micron sized substrate.
	
	\section*{References}
	\bibliographystyle{elsarticle-num.bst}
	\bibliography{references}

\begin{thebibliography}{10}
\expandafter\ifx\csname url\endcsname\relax
  \def\url#1{\texttt{#1}}\fi
\expandafter\ifx\csname urlprefix\endcsname\relax\def\urlprefix{URL }\fi
\expandafter\ifx\csname href\endcsname\relax
  \def\href#1#2{#2} \def\path#1{#1}\fi

\bibitem{grove1852vii}
W.~R. Grove, {On} the electro-chemical polarity of gases, Philosophical
  Transactions of the Royal Society of London~(142) (1852) 87--101.

\bibitem{faraday1857x}
M.~Faraday, {X}. {The} {Bakerian} {Lecture} - {Experimental} relations of gold
  (and other metals) to light, Philosophical Transactions of the Royal Society
  of London~(147) (1857) 145--181.

\bibitem{hawkeye2014glancing}
M.~M. Hawkeye, M.~T. Taschuk, M.~J. Brett, Glancing angle deposition of thin
  films: {Engineering} the nanoscale, John Wiley \& Sons, 2014.

\bibitem{nikitenkov2017modern}
N.~Nikitenkov, Modern Technologies for Creating the Thin-film Systems and
  Coatings, BoD--Books on Demand, 2017.

\bibitem{lakhtakia2005sculptured}
A.~Lakhtakia, R.~Messier, Sculptured thin films: Nanoengineered morphology and
  optics, Vol. 143, SPIE press, 2005.

\bibitem{mattox2003vacuum}
D.~M. Mattox, V.~Mattox, Vacuum coating technology, Springer, 2003.

\bibitem{martin2009handbook}
P.~M. Martin, Handbook of deposition technologies for films and coatings:
  {Science}, applications and technology, 3rd Edition, William Andrew, 2010.

\bibitem{zhao2003designing}
Y.~Zhao, D.~Ye, G.-C. Wang, T.-M. Lu, Designing nanostructures by glancing
  angle deposition, in: Nanotubes and Nanowires, Vol. 5219, International
  Society for Optics and Photonics, 2003, pp. 59--73.

\bibitem{Divakar}
S.~M. Haque, R.~De, A.~Mitra, J.~Misal, C.~Prathap, P.~V. Satyam, K.~D. Rao,
  Demonstration of tunable {Ag} morphology from nanocolumns to discrete
  nanoislands using novel angle constrained glancing angle {EB} evaporation
  technique, Surface and Coatings Technology 375 (2019) 363--369.

\bibitem{keller2011polycrystalline}
A.~Keller, L.~Peverini, J.~Grenzer, G.~J. Kovacs, A.~M{\"u}cklich, S.~Facsko,
  Polycrystalline ni thin films on nanopatterned si substrates: From highly
  conformal to nonconformal anisotropic growth, Physical Review B 84~(3) (2011)
  035423.

\bibitem{review2016}
A.~Barranco, A.~Borras, A.~R. Gonzalez-Elipe, A.~Palmero, Perspectives on
  oblique angle deposition of thin films: From fundamentals to devices,
  Progress in Materials Science 76 (2016) 59--153.

\bibitem{henderson1974simulation}
D.~Henderson, M.~Brodsky, P.~Chaudhari, Simulation of structural anisotropy and
  void formation in amorphous thin films, Applied Physics Letters 25~(11)
  (1974) 641--643.

\bibitem{dirks1977columnar}
A.~Dirks, H.~Leamy, Columnar microstructure in vapor-deposited thin films, Thin
  solid films 47~(3) (1977) 219--233.

\bibitem{karabacak2011thin}
T.~Karabacak, Thin-film growth dynamics with shadowing and re-emission effects,
  Journal of Nanophotonics 5~(1) (2011) 052501.

\bibitem{kim1977computer}
S.~Kim, J.~Henderson, P.~Chaudhari, Computer simulation of amorphous thin films
  of hard spheres, Thin Solid Films 47~(2) (1977) 155--158.

\bibitem{jones1967re}
R.~Jones, C.~Standley, L.~Maissel, Re-emission coefficients of ${Si}$ and
  ${SiO_2}$ films deposited through rf and dc sputtering, Journal of Applied
  Physics 38~(12) (1967) 4656--4662.

\bibitem{drotar2000mechanisms}
J.~T. Drotar, Y.-P. Zhao, T.-M. Lu, G.-C. Wang, Mechanisms for plasma and
  reactive ion etch-front roughening, Physical Review B 61~(4) (2000) 3012.

\bibitem{drotar2000surface}
J.~T. Drotar, Y.-P. Zhao, T.-M. Lu, G.-C. Wang, Surface roughening in shadowing
  growth and etching in 2+ 1 dimensions, Physical Review B 62~(3) (2000) 2118.

\bibitem{meakin1988ballistic}
P.~Meakin, Ballistic deposition onto inclined surfaces, Physical Review A
  38~(2) (1988) 994.

\bibitem{meakin1998fractals}
P.~Meakin, Fractals, scaling and growth far from equilibrium, Vol.~5, Cambridge
  university press, 1998.

\bibitem{garcia2018growth}
A.~Garcia-Valenzuela, R.~{\'A}lvarez, V.~Rico, J.~Cotrino, A.~R.
  Gonzalez-Elipe, A.~Palmero, Growth of nanocolumnar porous {$TiO_2$} thin
  films by magnetron sputtering using particle collimators, Surface and
  Coatings Technology 343 (2018) 172--177.

\bibitem{troncoso2020silver}
G.~Troncoso, J.~M. Garc{\'\i}a-Mart{\'\i}n, M.~Gonz{\'a}lez, C.~Morales,
  M.~Fern{\'a}ndez-Castro, J.~Soler-Morala, L.~Gal{\'a}n, L.~Soriano, Silver
  nanopillar coatings grown by glancing angle magnetron sputtering for reducing
  multipactor effect in spacecrafts, Applied Surface Science (2020) 146699.

\bibitem{haque2017glancing}
S.~M. Haque, K.~D. Rao, S.~Tripathi, R.~De, D.~Shinde, J.~Misal, C.~Prathap,
  M.~Kumar, T.~Som, U.~Deshpande, N.~Sahoo, Glancing angle deposition of
  {$SiO_2$} thin films using a novel collimated magnetron sputtering technique,
  Surface and Coatings Technology 319 (2017) 61--69.

\bibitem{meakin1988invited}
P.~Meakin, R.~Jullien, Invited paper, {Simple} ballistic deposition models for
  the formation of thin films, in: Modeling of Optical Thin Films, Vol. 821,
  International Society for Optics and Photonics, 1988, pp. 45--55.

\bibitem{mansour2019ballistic}
T.~Mansour, R.~Rastegar, A.~Roitershtein, On ballistic deposition process on a
  strip, Journal of statistical physics 177~(4) (2019) 626--650.

\bibitem{onlineprotractor}
Online protractor, https://www.ginifab.com/feeds/angle\_measurement/, last
  Accessed: 10 April, 2021.

\bibitem{tangentrule}
J.~Nieuwenhuizen, H.~Haanstra, Microfractography of thin films, Philips Tech
  Rev 27~(3) (1966) 87--91.

\bibitem{cosRule}
R.~Tait, T.~Smy, M.~Brett, Modelling and characterization of columnar growth in
  evaporated films, Thin Solid Films 226~(2) (1993) 196--201.

\bibitem{material1}
Y.~Zhao, Y.~He, C.~Brown, Composition dependent nanocolumn tilting angle during
  the oblique angle co-deposition, Applied Physics Letters 100~(3) (2012)
  033106.

\bibitem{material2}
H.~Zhu, W.~Cao, G.~K. Larsen, R.~Toole, Y.~Zhao, Tilting angle of nanocolumnar
  films fabricated by oblique angle deposition, Journal of Vacuum Science \&
  Technology B, Nanotechnology and Microelectronics: Materials, Processing,
  Measurement, and Phenomena 30~(3) (2012) 030606.

\bibitem{Lola}
G.~G. Lola, J.~Parra-Barranco, J.~R. S{\'a}nchez-Valencia, A.~Barranco,
  A.~Borr{\'a}s, A.~R. Gonz{\'a}lez-Elipe, M.~C. Garc{\'\i}a-Guti{\'e}rrez,
  J.~J. Hern{\'a}ndez, D.~R. Rueda, T.~A. Ezquerra, Correlation lengths,
  porosity and water adsorption in {$TiO_2$} thin films prepared by glancing
  angle deposition, Nanotechnology 23~(20) (2012) 205701.

\bibitem{Alvarez}
R.~Alvarez, L.~Gonz{\'a}lez-Garc{\'\i}a, P.~Romero-G{\'o}mez, V.~Rico,
  J.~Cotrino, A.~R. Gonz{\'a}lez-Elipe, A.~Palmero, Theoretical and
  experimental characterization of {$TiO_2$} thin films deposited at oblique
  angles, Journal of Physics D: Applied Physics 44~(38) (2011) 385302.

\bibitem{sood}
A.~W. Sood, D.~J. Poxson, F.~W. Mont, S.~Chhajed, J.~Cho, E.~F. Schubert, R.~E.
  Welser, N.~K. Dhar, A.~K. Sood, Experimental and theoretical study of the
  optical and electrical properties of nanostructured indium tin oxide
  fabricated by oblique-angle deposition, Journal of nanoscience and
  nanotechnology 12~(5) (2012) 3950--3953.

\bibitem{MD2013}
B.~C. Hubartt, X.~Liu, J.~G. Amar, Large-scale molecular dynamics simulations
  of glancing angle deposition, Journal of Applied Physics 114~(8) (2013)
  083517.

\bibitem{porDistr2020}
T.~Ott, G.~Gerlach, Morphological characterization and porosity profiles of
  tantalum glancing-angle-deposited thin films, Journal of Sensors and Sensor
  Systems 9~(1) (2020) 79--87.

\bibitem{meakin1987restructuring}
P.~Meakin, R.~Jullien, Restructuring effects in the rain model for random
  deposition, journal de Physique 48~(10) (1987) 1651--1662.

\bibitem{krug1989microstructure}
J.~Krug, P.~Meakin, Microstructure and surface scaling in ballistic deposition
  at oblique incidence, Physical Review A 40~(4) (1989) 2064.

\bibitem{tait1990ballistic}
R.~Tait, T.~Smy, M.~Brett, A ballistic deposition model for films evaporated
  over topography, Thin Solid Films 187~(2) (1990) 375--384.

\bibitem{banerjee2014surface}
K.~Banerjee, J.~Shamanna, S.~Ray, Surface morphology of a modified ballistic
  deposition model, Physical Review E 90~(2) (2014) 022111.

\bibitem{mal2016surface}
B.~Mal, S.~Ray, J.~Shamanna, Surface properties and scaling behavior of a
  generalized ballistic deposition model, Physical Review E 93~(2) (2016)
  022121.

\bibitem{luo}
Y.~Luo, M.~Lin, N.~Zhou, H.~Huang, C.-T. Tsai, L.~Zhou, Molecular dynamics
  simulation study of the microstructure of {a-Si}: H thin film grown by
  oblique-angle deposition, Physica B: Condensed Matter 545 (2018) 80--85.

\bibitem{bouaouina2018nanocolumnar}
B.~Bouaouina, C.~Mastail, A.~Besnard, R.~Mareus, F.~Nita, A.~Michel,
  G.~Abadias, Nanocolumnar tin thin film growth by oblique angle
  sputter-deposition: Experiments vs. simulations, Materials \& Design 160
  (2018) 338--349.

\end{thebibliography}
\end{document}